\documentclass[a4paper]{article}
\usepackage[section]{algorithm}
\usepackage{algorithmic}
\usepackage{a4wide,amsmath,amsthm,amsfonts}
\usepackage{graphicx}
\newtheorem{defi}{Definition}[section]
\newtheorem{theo}{Theorem}[section]
\newtheorem{prop}{Property}[section]
\newtheorem{cor}{Corollary}[section]
\newcommand{\GO}{{\cal O}}

\title{Computing the Kalman form}

\author{Cl\'ement Pernet \\ LMC, Universit\'e Joseph Fourier\\51, rue
  des Math\'ematiques BP 53 IMAG-LMC 38041 Grenoble, FRANCE\\
  \texttt{clement.pernet@imag.fr} \and 
  Aude Rondepierre\\ LMC, Universit\'e Joseph Fourier\\51, rue
  des Math\'ematiques BP 53 IMAG-LMC 38041 Grenoble, FRANCE\\
  \texttt{aude.rondepierre@imag.fr}
  \and Gilles  Villard\\ CNRS, LIP,
  Ecole Normale Sup\'erieure de Lyon\\46, All\'ee d'Italie, 69364 Lyon
  Cedex 07 FRANCE\\
  \texttt{gilles.villard@ens-lyon.fr}}

\begin{document}
\maketitle
\begin{abstract}
We present two algorithms for the computation of the Kalman form of
a linear control system. The first one is based on the technique
developed by Keller-Gehrig for the computation of the characteristic
polynomial. The cost is a logarithmic number of matrix
multiplications. To our knowledge, this improves the best previously
known algebraic complexity by an order of magnitude. Then we also
present a cubic algorithm proven to be more efficient in practice.
\end{abstract}

\section{Introduction}
This report is a continuation of a collaboration of the first two authors
on the algorithmic similarities between the computation of the Kalman form and 
of the characteristic polynomial. This collaboration led 
to \cite[Theorem 2]{DR:2005:kalman}. We report here an improvement of this last 
result based on a remark by the third author.

For a definition of the Kalman form of a linear control system, see
\cite[Theorem 1]{Kalman:1961}.

In this report we show how to adapt the branching algorithm of Keller-Gehrig
\cite[\S5]{Keller-Gehrig:1985}  (computing the characteristic
polynomial) to compute the Kalman form. This implies an
algebraic time complexity of $\GO(n^{\omega}\text{log}n)$.
Now, the discussion of \cite[\S2]{DPW:2005:charpoly} shows that a
cubic algorithm, \texttt{LUK}, is more efficient in practice for the
computation of the characteristic polynomial. Therefore, we adapt it
to the computation of the Kalman form.

The outline of this report is the following : in section \ref{sec:ckm} we define
the compressed Kyrlov matrix. It will help to describe the adaptation
of  Keller-Gehrig's
algorithm to the computation of the Kalman form.
In section \ref{sec:kg}, we recall Keller-Gehrig's algorithm. Section 
\ref{sec:kalman} presents the main result of this report, on the time complexity
of the computation of the Kalman form. Lastly, we give a full description
two algorithms to compute the Kalman form. The first one precises the operations
used in section \ref{sec:kalman} to achieve the complexity and improves
the constant hidden in the $\GO()$ by saving operations. 
The second is  based on another algorithm for
the characteristic polynomial, that does not achieve the same algebraic 
complexity, but appears to be faster in practice.

We will denote by $\omega$ the exponent in the complexity  of the matrix 
multiplication.

\section{The compressed Krylov matrix} \label{sec:ckm}

Let $A$ and $B$ be two matrices of dimension respectively $n\times n$ and $n \times m$.
Consider the $n\times (mn)$ Krylov matrix $K$ generated by the $m$
column vectors of $B$ and their iterates with the matrix $A$~:
$$
K = \left[ \begin{array}{c|c|c|c|c|c|c}
    b_1 & \dots & A^{n-1}b_1 & \dots & b_m & \dots & A^{n-1} b_m
  \end{array}\right]
$$

Let $r$ be the rank of $K$. $r \geq \text{rank}(B)$. Let us form the $n\times r$ 
non-singular matrix $\overline{K}$ by picking the first $r$ linearly independent 
columns of $K$.

\begin{defi}
$\overline{K}$ is the compressed Krylov matrix of $B$ relatively to $A$.
\end{defi}

If a column vector $A^kb_j$ is linearly dependent  with the
previous column vectors, then any vector $A^lb_j, l>k$ will also be linearly
dependent. Consequently the matrix $\overline{K}$ has the form :

\begin{equation}\label{eq:krylovbasis}
\overline{K} = \left[
  \begin{array}{c|c|c|c|c|c|c}
    b_1 & \dots & A^{d_1-1}b_1 & \dots & b_m & \dots & A^{d_m-1} b_m
  \end{array}
  \right]
\end{equation}
for some $d_i$ such that $0\leq d_i \leq n-1$ and $\sum_{i=1}^m{d_i}=r$.

The order in the choice of the independent column vectors (from the left to 
the right) can also be interpreted in terms of lexicographical order on the sequence
$(d_i)$. Following Storjohann \cite{Storjohann:2000:thesis}, we can therefore also
define the compressed Krylov matrix as follows~:

\begin{defi}
The compressed Krylov matrix of $B$ relatively to $A$ is a matrix of the form
$$
\left[
  \begin{array}{c|c|c|c|c|c|c}
    b_1 & \dots & A^{d_1-1}b_1 & \dots & b_m & \dots & A^{d_m-1} b_m
  \end{array}
  \right]
$$
of rank $r$, such that the sequence $(d_i)$ is lexicographically maximal.
\end{defi}

The next section will present an algorithm to compute this compressed Krylov matrix.

\section{Keller-Gehrig's algorithm} \label{sec:kg}

The selection of the linearly independent columns, starting from left to right,
can be done by a gaussian elimination. 
A block elimination is mandatory to reduce the algebraic complexity to
matrix multiplication.
For this task, Keller-Gehrig first introduced in \cite[\S4]{Keller-Gehrig:1985} an
algorithm called ``step form elimination''. The more recent
litterature replaced it by the row echelon elimination (for example in
\cite{Burgisser:1997}).
We showed in \cite{DPW:2005:charpoly} that the \texttt{LQUP}  elimination 
(defined in \cite{Ibarra:1982:LSP}) of $\overline{K}^T$ 
could also be used (algorithm \ref{alg:CRF}). This last algorithm simply
returns the submatrix formed by the first independent column vectors of
the input matrix form left to right.
\begin{algorithm}[phtb]
\caption{\texttt{ColReducedForm}}\label{alg:CRF}
\begin{algorithmic}[1]
\REQUIRE{$A$ a $m \times n$ matrix of rank $r$ ($m,n\geq r$) over a field}
\ENSURE{$A'$ a $m \times r$ matrix formed by r linearly independent
columns of $A$}
\STATE $(L,Q,U,P,r)=\text{LQUP}(A^T)$ ($r=rank(A)$)
\STATE return $([I_r 0](Q^TA^T))^T$
\end{algorithmic}
\end{algorithm}

Thus a straightforward algorithm to compute $\overline{K}$ would be to run
algorithm \ref{alg:CRF}
on the matrix $K$. But cost of the computation of $K$ is prohibitive 
($n^3$ coefficients and $\GO(n^4)$  arithmetic operations with standard
matrix product).
Hence, the elimination process must be combined within the building of the
matrix.

The computation of the iterates can rely on matrix multiplication, by
computing the  $\lceil \text{log}_2(n)\rceil$ following powers of $A$ : 
$$
A,A^2,\dots,A^{2^i},A^{2^{\lceil\text{log}_2(n)\rceil-1}}
$$
Thus the following scheme, 
\begin{equation}\label{eq:it}
\left\{
\begin{array}{lcl}
V_0 &=& [b_j]\\
V_{i+1} &=& [V_i | A^{2^{i}} V_i]
\end{array}
\right.
\end{equation} 
where the matrix $V_i$ has $2^i$ columns, computes every iterates of $b_j$ in 
$\GO(n^\omega\text{log}n)$ operations.

One elimination is performed after each application of $A^{2^i}$,
to discard the linearly dependent iterates for the next iteration
step.
Moreover if a vector $b_j$ has only $k<2^i$ linearly independent iterates, one
can stop the computation of its iterates. Therefore, the scheme
(\ref{eq:it}) will only be applied on the block iterates of size $2^i$.

From these remarks, we can now present Keller-Gehrig's algorithm.
Although is was initially designed for the computation of the characteristic 
polynomial, we prefer to show it in a more general setting~: the computation
of the compressed Krylov matrix. Afterwards, we will show that the computation of 
the characteristic polynomial is a specialization of this algorithm with $B=I_n$ and
that the recover of its coefficients is straightforward.

\begin{algorithm}[htbp]
\caption{Compressed Krylov Matrix [Keller-Gehrig]}\label{alg:kgckm}
\begin{algorithmic}[1]
\REQUIRE{$A$ a $n \times n$ matrix over a field, $B$, a $n\times m$ matrix}
\ENSURE{$(\overline{K},r)$} as in (\ref{eq:krylovbasis})
\STATE $i=0$
\STATE $V_0 = B = (V_{0,1},V_{0,2},\dots,V_{0,m})$
\STATE $ C=A $
\WHILE{($\exists k, V_k$ has $2^i$ columns)}
 \FORALL j
  \IF{ ( $V_{i,j}$ has strictly less than $2^i$ columns )}
  \STATE $W_j = V_{i,j}$
  \ELSE
  \STATE $W_j = \left[ V_{i,j} |CV_{i,j}\right]$
  \ENDIF
 \ENDFOR
 \STATE $W = [W_1|\dots|W_n]$
 \STATE $V_{i+1} = \text{ColReducedForm}( W )$ remember $r=rank(W)$
 \COMMENT{$V_{i+1} = [V_{i+1,1}|\dots|V_{i+1,n}]$ where $V_{i+1,j}$ are the remaining vectors of $W_j$ in $V_{i+1}$}
% \STATE keep track of the first linearly dependent row $l_j$ in $LQ$ corresponding to 
%the block $W_j$.
 \STATE $C = C \times C$
 \STATE $i=i+1$
\ENDWHILE
\STATE return $(V_i,r)$
\end{algorithmic}
\end{algorithm}

\begin{theo}[Keller-Gehrig] \label{th:kg}
Suppose $m = \GO(n)$. The compressed Krylov matrix of $B$ relatively to $A$
can be computed in $\GO(n^\omega\text{log}n)$ field operations.
\end{theo}

\begin{proof}
Algorithm \ref{alg:kgckm} satisfies the statement (cf \cite{Keller-Gehrig:1985}).
\end{proof}

\begin{prop}
Let $\overline{K}$ be the compressed Krylov matrix of the identity matix relatively to $A$.
The matrix $\overline{K}^{-1}A\overline{K}$ has the Hessenberg
polycyclic form : it is block upper triangular, with companion
blocks on its diagonal, and the upper blocks are zero except on their
last column.
\begin{small}
\begin{equation}\label{eq:hessenberg}
  \overline{K}^{-1}A\overline{K}= 
  \left[
    \begin{array}{rrr}
      \fbox{$
	\begin{array}{cccc}
	0 & & & *\\
	1 &0& & *\\
	  &\ddots&\ddots&*\\
	  &      & 1 &*
      \end{array}
	$}
       &&
      \begin{array}{cccc}
	& & & *\\
	& & & *\\
	& & &*\\
	& & &*
      \end{array}
      \\
        &\ddots& \\
      &&
      \fbox{$
	\begin{array}{cccc}
	0 & & & *\\
	1 &0& & *\\
	  &\ddots&\ddots&*\\
	  &      & 1 &*
      \end{array}
      $}
    \end{array}
%%     \begin{array}{cccc|cccc|c|cccc}
%%       0&      &      &  *    &     &      &      &  *   &      &     &      &      &  *   \\
%%       1&  0   &      &  *    &     &      &      &  *   &      &     &      &      &  *   \\
%%       &\ddots&\ddots&\vdots &     &      &      &\vdots&      &     &      &      &\vdots\\
%%       &      &   1  &  *    &     &      &      &  *   &      &     &      &      &  *   \\
%%       \hline
%%       &      &      &      &  0  &      &      &  *   &      &      &      &      &  *   \\ 
%%       &      &      &      &  1  &  0   &      &  *   &      &      &      &      &  *   \\ 
%%       &      &      &      &     &\ddots&\ddots&\vdots&      &      &      &      &\vdots\\
%%       &      &      &      &     &      &   1  &  *   &      &      &      &      &  *   \\ 
%%       \hline
%%       &      &      &      &     &      &      &      &\ddots&      &      &      &\vdots\\
%%       \hline
%%       &      &      &      &     &      &      &      &      &   0  &      &      &  *   \\
%%       &      &      &      &     &      &      &      &      &   1  &  0   &      &  *   \\
%%       &      &      &      &     &      &      &      &      &      &\ddots&\ddots&\vdots\\
%%       &      &      &      &     &      &      &      &      &      &      &   1  &  *   \\
%%     \end{array}
    \right]
\end{equation}
\end{small}
\end{prop}

\begin{cor}[Keller-Gehrig]
The characteristic polynomial of $A$ can be computed in $\GO(n^\omega\text{log}n)$
field operation.
\end{cor}
\begin{proof}
The characteristic polynomial of the shifted Hessenberg form (\ref{eq:hessenberg})
is the product of the polynomials associated to the companion blocks on its diagonal.
And since determinants are invariants under similarity transformations, it equals 
the characteristic polynomial of $A$.
\end{proof}

\section{Computation of the Kalman form} \label{sec:kalman}

Theorem \ref{th:kalmandef} recalls the definition of the Kalman form
of two matrices $A$ and $B$.

\begin{theo}\label{th:kalmandef}
Let $A$ and $B$ be two matrices of dimension respectively $n\times n$ 
and $n\times m$. Let $r$ be the dimension of $\text{Span}(B, AB, \dots, A^{n-1}B)$.
There exist a non singular matrix $T$ of dimension $n\times n$ such that
$$
T^{-1}AT = \left[\begin{array}{cc}H&X\\0&Y\end{array}\right], 
\left[\begin{array}{c} B_1\\0\end{array}\right] = T^{-1}B
$$
where $H$ and $B_1$ are respectively $r\times r$ et $r\times m$.
\end{theo}

The main result of this report is the following result, based on an idea by 
the third author.

\begin{theo}\label{th:kalman}
Let $V$ be compressed Krylov matrix of $B$ respectively to $A$. 
Complete $V$ into a basis $T$ of $K^n$ by adding $n-r$ columns at the end of $V$.
Then $T$ satisfies the definition of the Kalman form of $A$ and $B$.
\end{theo}

\begin{proof}
The matrix $V$ satisfy the relation  $$AV=VH$$ where $H$ is $r\times r$ and
has the Hessenberg polycylic form (\ref{eq:hessenberg}). 
Let us note $T=[V|W]$.

Now 
\begin{eqnarray*}
AT & = & \left[\begin{array}{c|c}AV&AW\end{array}\right]\\
& = &T\left[\begin{matrix}H&X\\0&Y\end{matrix}\right].
\end{eqnarray*}

Lastly, $V$ is a basis of $\text{Span}(B,AB,\dots,A^nB)$. Therefore each column of 
$B$ is a linear combination of the colmuns of $V$~:
$$
B = T \left[\begin{array}{c} B_1\\0\end{array}\right].
$$
\end{proof}

\begin{cor}
The Kalman form of $A$ and $B$ can be computed in $\GO(n^\omega\text{log}n)$.
\end{cor}

\begin{proof}
Applying theorem \ref{th:kg}, there only remains to show how to complete $V$
into $T$ in $\GO(n^\omega\text{log}n)$.
The idea is to complete $V$ in its triangularized form. One computes the
LUP factorization of $V^T$~: 
$$
V^T = [L][\begin{array}{cc}U_1&U_2\end{array}]P
$$
Then replace $[U_1 U_2]$ by
$\left[\begin{array}{cc}U_1 &U_2\\0&Id\end{array}\right]$ and 
$[L]$ by $\left[\begin{array}{cc}L&0\\0&Id\end{array}\right]$ to get a $n\times n$
non singular matrix.
This simply corresponds to set
$$
T=\left[\begin{array}{c|c}\overline{K} & P^T\left[\begin{array}{c}0\\I_{n-r}\end{array}\right]\end{array} \right].
$$

It only costs $\GO(n^\omega)$ field operations to recover the whole Kalman
form (blocks $H,X,Y$ and $B_1$), using for example matrix multiplications
and matrix inversions. See section \ref{sec:kalmankg} for more details.
\end{proof}

This last result improves the algebraic time complexity for
the computation of the Kalman form given 
in \cite[Theorem 2]{DR:2005:kalman} by an order of
magnitude.

\section{Algorithms into practice}

The goal of the previous section was to establish the time complexity estimate and
we therefore only sketched the algorithms involved. We will now focus more precisely
on the operations so as to reduce the consant hiden in the $\GO()$ notation.

\subsection{Improvements on Keller-Gehrig's algorithm} \label{sec:minpoly}

The first improvement concerns the recover of the Hessenberg polycyclic form 
\ref{eq:hessenberg}, once the compressed Krylov matrix is computed. In 
\cite{Keller-Gehrig:1985} Keller-Gehrig simply suggests to compute the product 
$K^{-1}AK$. This implies $4.66n^3$ additional field operation. We propose here 
to reduce this cost to $\phi n^2$, where $\phi$ is the number of blocks in
the Hessenberg form. This technique was presented in \cite{DPW:2005:charpoly}.
We recall and extend it here for the recovery of the whole Hessenberg 
polycylic form.

First consider the case where the $n$ first iterates of only one vector $v$ 
are linearly independent.

Let $K=[v|Av|\dots|A^{n}v]$. The last column is the first which is 
linearly dependent with the previous.
Let $P(X) = X^n - \sum_{i=0}^{n-1}m_iX^i$ represent this dependency (the minimal 
polynomial of this vector relatively to $A$).
Again consider the LUP factorization of $K^T$.
Let $X_{n+1}$ denote the $n+1$th row of the matrix $X$ and $X_{1\dots n}$ be the 
block of the first $n$ rows of $X$.
Then we have
$$
K^T_{n+1} = (A^nv)^T = (\sum_{i=0}^{n-1}m_iA^iv)^T = [\begin{array}{ccc}m_0&\dots&m_{n-1}\end{array}](K^T)_{1\dots n}
$$
Therefore
$$
L_{n+1} = [\begin{array}{ccc}m_0&\dots&m_{n-1}\end{array}]L_{1\dots n}.
$$

And the coefficients $m_i$ can be recovered as the solution of a triangular 
system.

Now, one easily check that 
$$
K_{1\dots n}^{-1}AK_{1\dots n} = 
\left[\begin{array}{cccc}
0 & & & m_0\\
1 &0& & m_1\\
  &\ddots&\ddots&\vdots\\
 &    & 1 & m_{n-1}
\end{array}\right].
$$

This companion matrix is the Hessenberg polycyclic matrix to be computed.

In the situation of Keller-Gehrig's algorithm, the linear depencies also involve
iterates of other vectors. However, the LQUP factorization will play a similar
role than the previous LUP and makes it possible to recover the whole
vector coefficients  of the linear dependency.

\begin{figure}[htbp]
\begin{center}
\includegraphics[width=.7\textwidth]{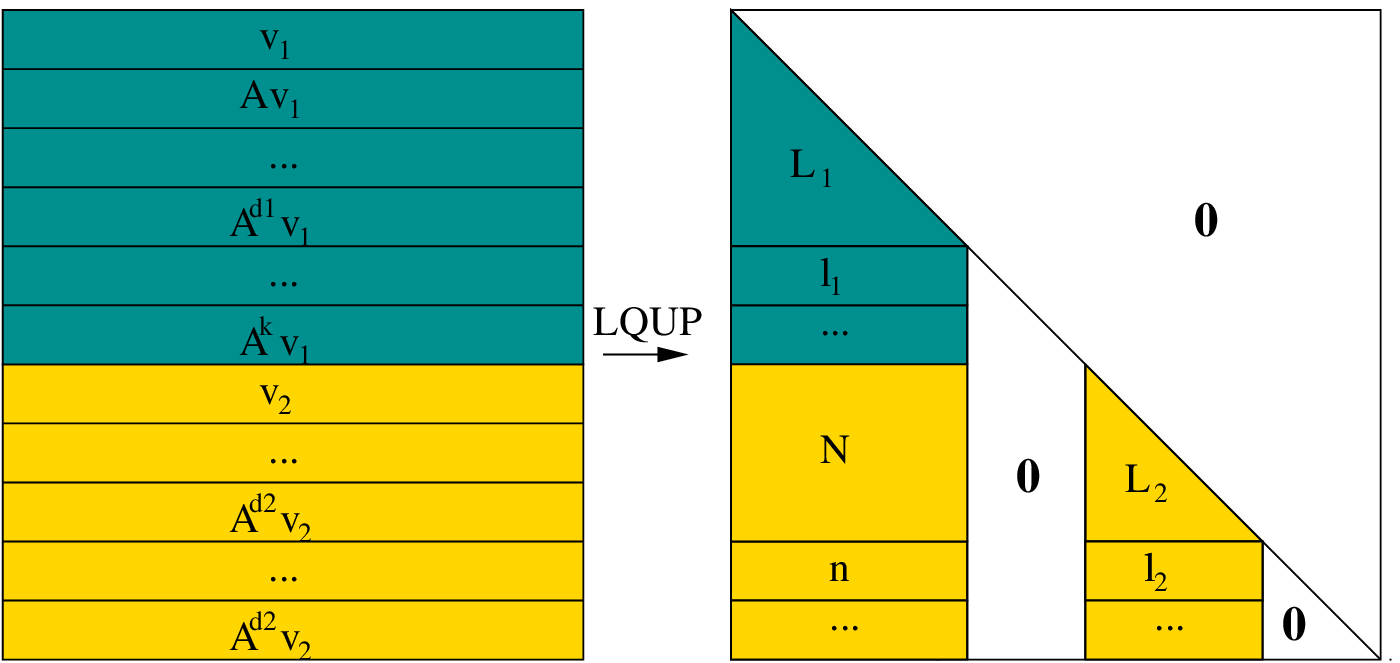}
\end{center}
\caption{LQUP factorization of 2 blocks of iterates}\label{fig:kgminpoly}
\end{figure}

\begin{figure}[htbp]
\begin{center}
\includegraphics[width=.6\textwidth]{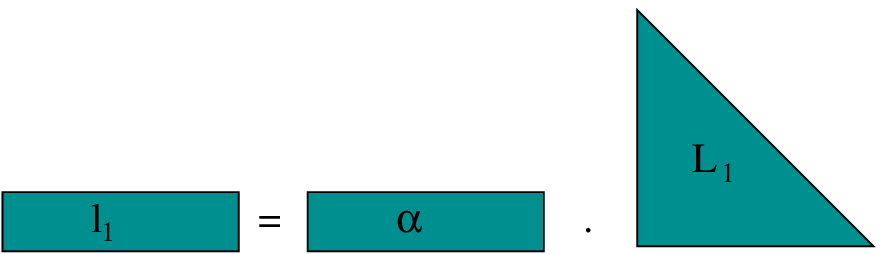}
\end{center}
\caption{Recover of the coefficients of the first linear dependency}\label{fig:kgminpolysysun}
\end{figure}

We show in figure \ref{fig:kgminpoly} the case of two blocks of iterates.
The first linear dependency relation (for $A^{d_1}v_1$) is done as previously 
(see figure \ref{fig:kgminpolysysun}).

Now for the second block, the first linearly dependent vector $A^{d_2}v_2$
satisfies a relation of the type~:
$$
A^{d_2}v_2 = \sum_{i=0}^{d_2-1}\beta_iA^iv_2 + \sum_{i=0}^{d_1-1}\gamma_iA^iv_1 
$$

\begin{figure}[htbp]
\begin{center}
\includegraphics[width=.6\textwidth]{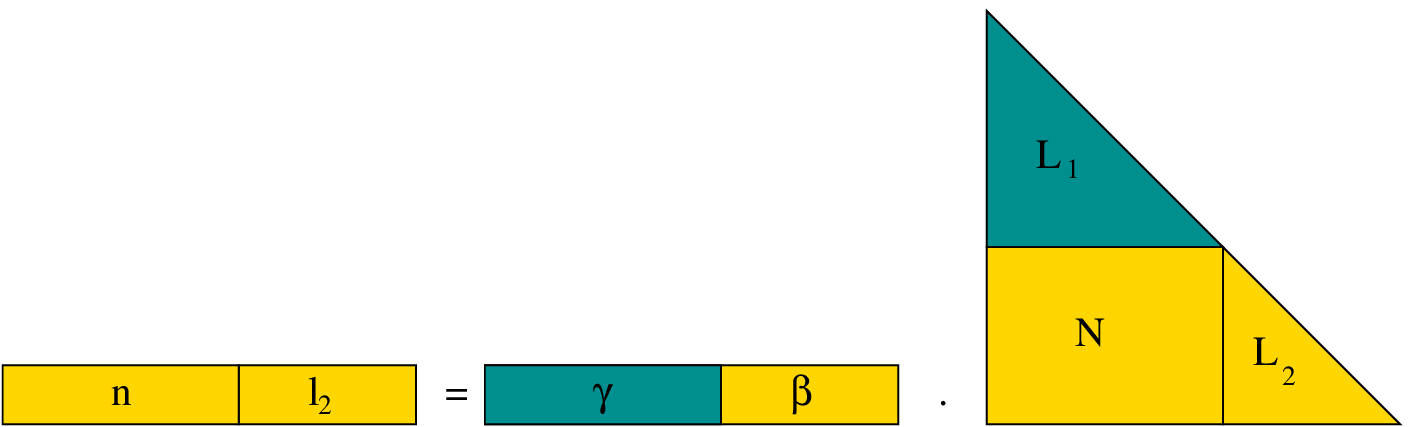}
\end{center}
\caption{Recover of the coefficients of the second linear dependency}\label{fig:kgminpolysys}
\end{figure}

The vector of coefficients $\beta=[\beta_i]$ and $\gamma=[\gamma_i]$ can
be obtained by solving the following system shown in figure \ref{fig:kgminpolysys}.

There only remains to build the Hessenberg polycyclic matrix from these vectors~:

$$
H=
\left[\begin{array}{cccccccc}
0 &      &      &\alpha_0      & &      &      &\gamma_0\\
1 &0     &      &\alpha_1      & &      &      &\gamma_1\\
  &\ddots&\ddots&\vdots        & &      &      & \vdots\\
  &      & 1    &\alpha_{d_1-1}& &      &      &\gamma_{d_1-1}\\
  &      &      &              &0&      &      &\beta_0       \\
  &      &      &              &1&0     &      &\beta_1       \\
  &      &      &              & &\ddots&\ddots&\vdots        \\
  &      &      &              & &      &1      &\beta_{d_2-1} \\
\end{array}\right].
$$

This technique can be applied to every block of iterates. Therefore the Hessenberg polycyclic matrix 
can be recovered by as many triangular system resolutions as the number of blocks.

\subsection{The main algorithm} \label{sec:kalmankg}

We have seen in section \ref{sec:kalman} how to compute the matrix $T$ (simply 
$T=\left[\begin{array}{c|c}\overline{K} & P^T\left[\begin{array}{c}0\\I_{n-r}\end{array}\right]\end{array} \right]$).
Section \ref{sec:minpoly} showed how to compute $H$. There only remains to 
show how to compute the matrices $X, Y$ and $B_1$ and we will be done.

We recall (from the proof of theorem \ref{th:kalman}) that
\begin{eqnarray*}
AT & = & \left[\begin{array}{c|c}AV&AP^T\left[\begin{matrix}0\\I_{n-r}\end{matrix}\right] \end{array}\right]\\
& = &T\left[\begin{matrix}H&X\\0&Y\end{matrix}\right]\\
\end{eqnarray*}
Then $X$ and $Y$ satisfy
$
T\left[\begin{matrix}X \\Y\end{matrix}\right]= 
AP^T\left[\begin{matrix}0 \\I_{n-r} \end{matrix}\right]
$. 
Let us write
$$
A'=PAP^T=\left[\begin{matrix}A'_{11}&A'_{12}\\A'_{21}&A'_{22}\end{matrix}\right].
$$
We have
\begin{equation*}
  T\left[\begin{matrix}X \\Y\end{matrix}\right]= 
  P^TA'\left[\begin{matrix}0 \\I_{n-r} \end{matrix}\right]=
  P^T\left[\begin{matrix}A'_{12}\\A'_{22}\end{matrix}\right]
\end{equation*}
Now 
$$
T = P^T 
\left[\begin{matrix}U_1^T&0\\U_2^T&I_{n-r}\end{matrix}\right]  
\left[\begin{matrix}L^T&0\\0&I_{n-r}\end{matrix}\right],
$$ 
therefore 
\begin{eqnarray*}
\left[\begin{matrix}U_1^T&0\\U_2&I_{n-r}\end{matrix}\right] 
\left[\begin{matrix}L^T&0\\0&I_{n-r}\end{matrix}\right]
\left[\begin{matrix}X \\ Y\end{matrix}\right] &=&
\left[\begin{matrix}A'_{12} \\ A'_{22}\end{matrix}\right]\\
\left[\begin{matrix}U_1^T&0\\U_2^T&I_{n-r}\end{matrix}\right] 
\left[\begin{matrix}L^TX\\ Y\end{matrix}\right]&=& 
\left[\begin{matrix}A'_{12} \\ A'_{22}\end{matrix}\right]\\
\end{eqnarray*}
And the system 
\begin{equation*}
\left\{
\begin{array}{lcl}
U_1^TL^TX&=&A'_{12}\\
U_2^TL^TX+Y&=&A'_{22}
\end{array}
\right.
\end{equation*}
has the following solution
\begin{equation*}
\left\{
\begin{array}{lcl}
X&=&L^{-T}U_1^{-T}A'_{12}\\
Y&=&A'_{22} - U_2^TU_1^{-T}A'_{12}
\end{array}
\right.
\end{equation*}

The computation of $B_1$ is straightforward from the equation
$TB_1=B$~: 
$$B_1 =  L^{-T}U_1^{-T}PB.$$

We are now able to write the algorithm.

\begin{algorithm}[phtb]
\caption{Kalman form}\label{alg:kalman}
\begin{algorithmic}[1]
\REQUIRE{$A$ a $n \times n$ matrix over a field, $B$, a $n\times m$ matrix}
\ENSURE{$r,T,H,X,Y,B_1$} as in theorem \ref{th:kalmandef}

\STATE $(V,r)=\texttt{CompressedKrylovMatrix}(A,B)$

\IF{ (r=n)}
\STATE return $(n,Id,A,\emptyset,\emptyset,B)$
\ELSE
\STATE $(L,[U_1 U_2],P)=\text{LUP}(V^T)$
\STATE  $T=\left[\begin{array}{c|c}V & P^T\left[\begin{array}{c}0\\I_{n-r}\end{array}\right]\end{array} \right]$
\STATE $B_1 = L^{-T}U_1^{-T}PB$
\STATE $A' = PAP^T = \left[\begin{array}{cc}A'_{11}&A'_{12}\\A'_{21}&A'_{22}\end{array}\right]$
\STATE $X = L^{-T}U_1^{-T}A'_{12}$
\STATE $Y = A'_{22} - U_2^TU_1^{-T}A'_{12}$
\FORALL j
 \STATE $m_j = l_j L^{-1}_j$ as explained in section \ref{sec:minpoly}
\ENDFOR
\STATE Build the polycyclic matrix $H$ using the $m_j$ as shown in section  \ref{sec:minpoly}.
\STATE return $(r,T,H,X,Y,B_1)$ 
\ENDIF
\end{algorithmic}
\end{algorithm}

Lastly, note that the LUP factorization of $V^T$ is already computed at the end 
of the call to \texttt{CompressedKrylovMatrix}. Thus, step 5 in 
algorithm \ref{alg:kalman} can be skipped.

\subsection{LU-Krylov~: a cubic variant}

In \cite{DPW:2005:charpoly}, we introduce an algorithm for the computation of the 
characteristic polynomial~: \texttt{LUK}. Alike Keller-Gehrig's algorithm, it
is also based on the Krylov iterates of several vectors and relies as much as possible 
on matrix multiplication. But the krylov iterates are computed with matrix vector 
products, so as to avoid the $\text{log}n$ factor in the time complexity. 
As a consequence it is $\GO(n^3)$ algorithm, but we showed that it was faster in 
practice.

Algorithm \ref{alg:kalmanluk} shows how to adapt this algorithm to the computation
of the Kalman form. We expect this algorithm to be the more efficient in practice.

\begin{algorithm}[phtb]
\caption{\texttt{Kalman-LUK} : Kalman-LU-Krylov}\label{alg:kalmanluk}
\begin{algorithmic}[1]
\REQUIRE{$A$ a $n \times n$ matrix over a field, $B$, a $n\times m$ matrix}
\ENSURE{$r, T, H, X, Y, B_1$} as in theorem \ref{th:kalman}
\STATE $v=B_1$
 \STATE $\left\{\begin{array}{l} K = \left[\begin{matrix}v&Av&A^2v&\dots\end{matrix}\right]\\
                          (L,[U_1|U_2],P) = \text{\texttt{LUP}}(K^T), r_1 = \mbox{rank}(K)\end{array}\right.$
\COMMENT{The matrix $K$ is computed on the fly :  at most $2r_1$ columns are computed}
\STATE $m = (m_1,\dots,m_{r1}) = L_{r_1+1}L_{1\dots r_1}^{-1}$
\STATE $f = X^{r_1}-\sum_{i=1}^{r_1}{m_iX^{i-1}}$
\IF { $(r_1 = n)$ }
\STATE return $(n,Id,A,\emptyset,\emptyset,B)$
\ELSE 
 \STATE $A' = PAP^T =
 \left[\begin{array}{cc}A'_{11}&A'_{12}\\A'_{21}&A'_{22}\end{array}\right]$
 where $A'_{11}$ is $r_1\times r_1$.
 \STATE $A_R = A'_{22}-U_2^TU_1^{-T}A'_{12}$
 \STATE $B' = \left[\begin{matrix}L^{-T}&0\\0&I\end{matrix}\right]
\left[\begin{matrix}U_1^{-T}&0\\-U_2^TU_1^{-T}&I\end{matrix}\right]P B$
 \STATE Compute the permutation $Q$ s.t. $B'Q=\left[\begin{matrix}*&*\\0&Z\end{matrix}\right]$
 \COMMENT{$Z$ is $(n-r_1) \times \mu$}
 \IF{$(\mu = 0)$}
  \STATE $X = L^{-T}U_1^{-T}A'_{12}$
  \STATE $Y = A_R$
  \STATE $T = \left[\begin{array}{c|c}K&P^T\left[\begin{matrix}0\\I_{n-r_1}\end{matrix}\right]\end{array}\right]$
  \STATE return $(r_1,T,C_{f},X,Y,X)$
 \ELSE
  \STATE $(r_2,T^{(2)},H^{(2)},X^{(2)},Y^{(2)},B_1^{(2)}) = \texttt{Kalman-LUK} ( A_R, Z )$
 \STATE $T = \left[\begin{array}{c|c}K&P^T\left[\begin{matrix}0\\T^{(2)}\end{matrix}\right]\end{array}\right]$
 \STATE $J = L^{-T}U_1^{-T}A'_{12}T^{(2)} = \left[\begin{matrix}J_1&J_2\end{matrix}\right]$ 
 \COMMENT{$J_1$ is $r_1 \times r_2$ and $J_2$, $r_1\times (n-r_1-r_2)$}
 \STATE $H = \left[\begin{matrix}C_{f}&J_1\\0&H^{(2)}\end{matrix}\right]$
 \STATE $X = \left[\begin{matrix}J_2\\X^{(2)}\end{matrix}\right]$
 \STATE return $(r_1+r_2,T,H,X,Y^{(2)},B_1)$
 \ENDIF
\ENDIF
\end{algorithmic}
\end{algorithm}

%\begin{small}
\bibliographystyle{plain}
\bibliography{kalmankg}  
%\end{small}
\end{document}